\documentclass[aps,prl,twocolumn,showpacs,superscriptaddress,groupedaddress]{revtex4} 
\usepackage{SIunits}
\usepackage{float}
\usepackage{graphicx}
\usepackage{dcolumn}
\usepackage{bm}

\begin{document}

\title{Strain-Induced Hybrid Improper Ferroelectricity in Simple Perovskites from First Principles}
\author{Qibin Zhou and Karin M. Rabe}
\address{Department of Physics and Astronomy, Rutgers University, Piscataway, New Jersey 08855-0849}
\date{\today}
\begin{abstract}
We performed first principles calculations for epitaxially strained orthorhombic CaTiO$_3$. The computational results reveal the existence of a metastable ferroelectric phase at compressive strain with unexpected in-plane polarization. Symmetry analysis indicates that a strain-induced nonpolar instability at the X point combines with distortion modes present in the nonpolar orthorhombic structure to induce the polarization, leading to characterization as a hybrid improper ferroelectric. Our work demonstrates a mechanism for improper ferroelectricity in simple perovskites without the need for symmetry lowering by layering or cation ordering. 
\end{abstract}
\maketitle

Recently, improper ferroelectricity~\cite{PhysRevB.72.100103, cheong2007multiferroics, bousquet2008improper, picozzi2009first, PhysRevB.79.094416,Benedek201211, fukushima2011large, PhysRevLett.106.257601, stroppa2011electric, stroppa2013hybrid} has been the subject of intensive theoretical and experimental investigation. The difference between proper and improper ferroelectricity lies in the driving factor that leads to electric polarization. Unlike proper ferroelectrics, in which an unstable polar lattice distortion is the driving factor, the driving factor in improper ferroelectricity is a nonpolar symmetry-breaking ordering, such as a magnetic ordering~\cite{PhysRevB.79.094416, PhysRevLett.106.257601} or a nonpolar structural distortion~\cite{bousquet2008improper, Benedek201211,fukushima2011large, stroppa2011electric, stroppa2013hybrid}, which induces the polarization as a secondary order parameter. The difference in the underlying mechanism gives improper ferroelectrics some features distinct from proper ferroelectrics. For instance, in improper ferroelectric perovskites $AB$O$_3$, a $B$-cation with empty $d$-state is no longer required  as in most proper ferroelectric perovskites~\cite{cohen1992origin}, which allows the coexistence of magnetic order and polarization and makes improper ferroelectrics good candidates for new multiferroics. 

In improper ferroelectrics, polarization is induced by the nonpolar distortion through a bivariate term  $P\Phi^\gamma$, in which $P$ is polarization and $\Phi$ is the nonpolar ordering. With any non-zero value for $\Phi$, the energy can always be lowered by some nonzero value of the polarization in an appropriate direction. Recently, hybrid improper ferroelectricity~\cite{benedek2011hybrid}, as a generalization of improper ferroelectricity, has attracted increasing interest. In hybrid improper ferroelectrics, polarization is coupled with a combination of two nonpolar orderings through an interaction term of the form $P\Phi_1\Phi_2$, which is allowed if $\Phi_1$ and $\Phi_2$ combine to break all the symmetries forbidding polarization in the high-symmetry reference state. 
Due to the layered structure of perovskite superlattices~\cite{bousquet2008improper} and perovskite Ruddlesden-Popper phases~\cite{benedek2011hybrid}, the symmetry of the reference state is low enough so that two oxygen octahedron rotation modes can combine to remove the remaining inversion centers and induce a nonzero polarization.  However, in simple perovskites, the symmetry of the ideal perovskite reference state is too high for hybrid improper ferroelectricity based on oxygen-octahedron rotations, which preserve the inversion symmetry at the $B$-site. 

In this paper, we present a mechanism for hybrid improper ferroelectricity in a simple perovskite system, and identify and investigate a materials realization in epitaxially strained CaTiO$_3$ using a first-principles approach. The hybrid improper ferroelectric phase is a metastable P2$_1$ phase with in-plane polarization, unexpected for compressively strained CaTiO$_3$.  Through a detailed symmetry analysis, we show that the instability in the zone-boundary mode $X_5^-$ induced by compressive strain combines with the distortions which already present in the ground state $Pbnm$ structure to induce the in-plane ferroelectric polarization.

In this study, the first-principles calculations were performed using density functional theory within the local density approximation as implemented in VASP5.2~\cite{PhysRevB.47.558,PhysRevB.54.1169}. We used the Ceperly and Alder functional~\cite{PhysRevLett.45.566}, with projector-augmented wave potentials~\cite{PhysRevB.50.17953,PhysRevB.59.1758}. We considered  3p and 4s as valence states to build the Ca pseudopotential, 3p, 3d and 4s valence states for the Ti pseudopotential and 2p valence states
for the O pseudopotential. A plane-wave energy cut-off of 680 eV was used. The Brillouin zone of the twenty-atom unit cell was sampled by a $8\times8\times6$ Monkhorst-Pack $k$-point mesh~\cite{PhysRevB.13.5188}.

\begin{figure}
\includegraphics[width=3.0 in]{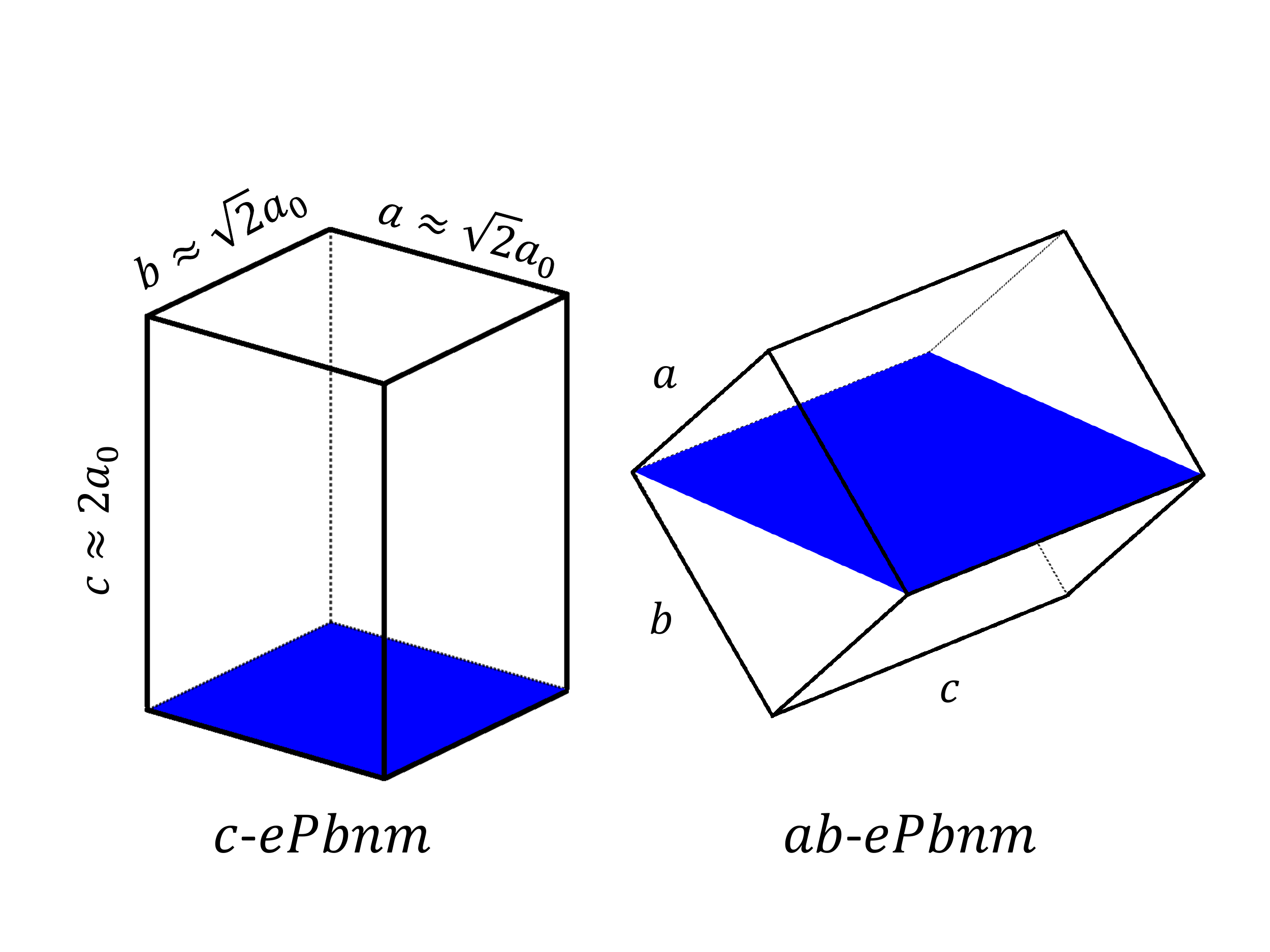}
\caption{The two ways to fit the $Pbnm$ structure of CaTiO$_3$ to a square $(001)$ substrate. The blue plane represents matching plane.}
\label{fig:strainApplication}
\end{figure}

To investigate the effect of epitaxial strain, we performed ``strained bulk" calculations for periodic crystals~\cite{PhysRevB.74.094104}, with appropriate constraints imposed on the lattice vectors. The zero-strain lattice constant is taken as $a_0=3.77$\AA, the cube root of the volume per formula unit of the relaxed $Pbnm$ structure. As shown in Fig.~\ref{fig:strainApplication}, there are two ways to fit the $Pbnm$ structure to the matching plane, which lead to two different epitaxially strained structures, $c$-$ePbnm$ ($[001]$) and $ab$-$ePbnm$ ($[110]$)~\cite{PhysRevB.79.220101}, respectively. In $c$-$ePbnm$, the $Pbnm$ space group is retained, while in $ab$-$ePbnm$, the symmetry is lowered to $P2_1/m$. In the structural optimization, both the lattice and internal structural parameters were relaxed until forces on atoms were less than 1 meV/\AA. Zone center phonons were obtained from frozen phonon calculations carried out in the relaxed non-polar structures, with each atom displaced by 0.01 \AA.

\begin{figure}
\includegraphics[width=3.7 in]{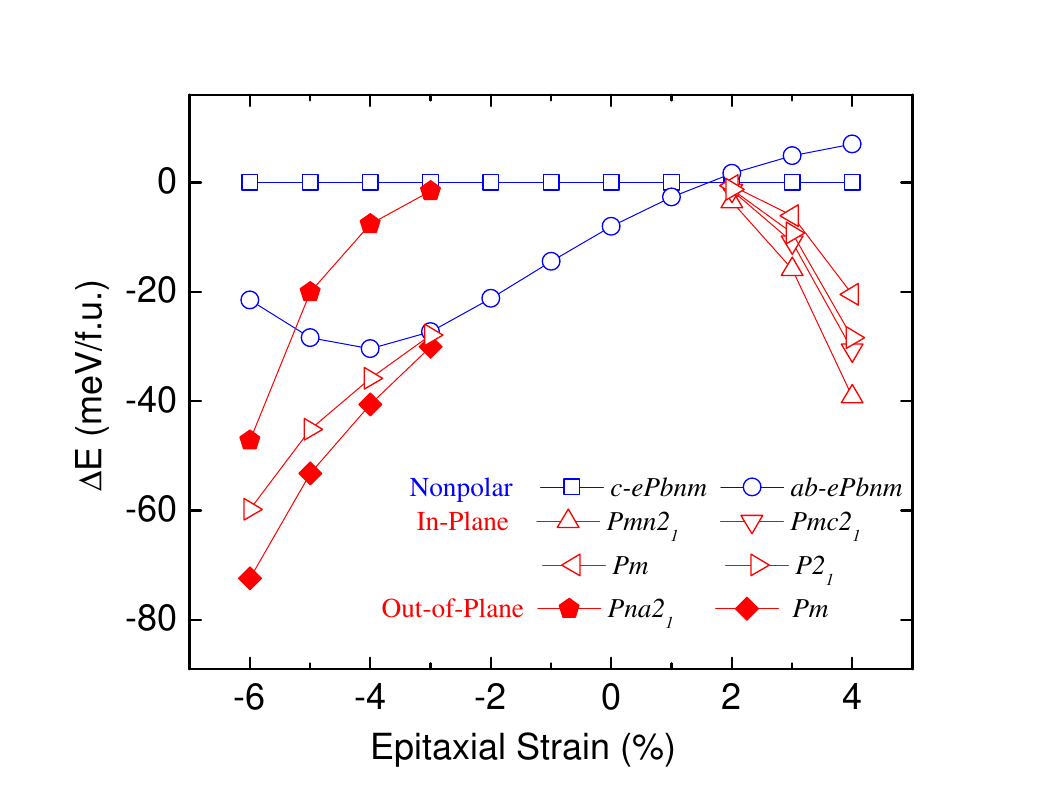}
\caption{Total energy per formula unit for various epitaxially constrained structures as a function of square misfit strain. At each strain, the energy of $c$-$ePbnm$ structure is taken as the zero of energy.}
\label{fig:first-principles energy}
\end{figure}
\begin{figure}
\includegraphics[width=3.7 in]{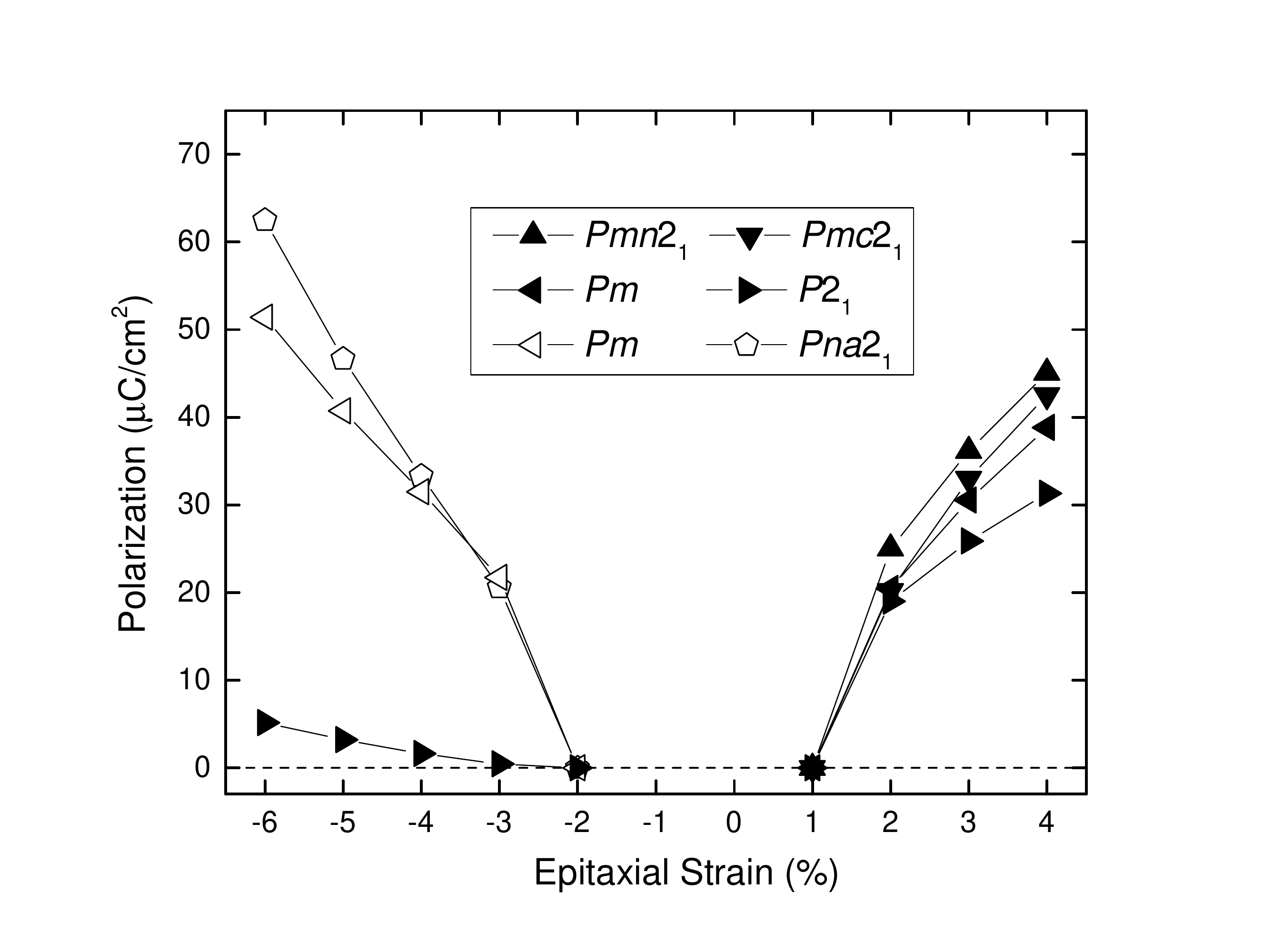}
\caption{First-principles calculated polarization in various polar structures.}
\label{fig:polarization}
\end{figure}

The energies of various structures are plotted as a function of epitaxial strain in Fig.~\ref{fig:first-principles energy}. Consistent with previous work~\cite{PhysRevB.79.220101}, the two non-polar structures show different trends with epitaxial strain, with an orientational phase transition point at approximately +1.5\%. Above +1.5 \% tensile strain, four polar structures with in-plane polarization were observed: $Pmn2_1$ ($[100]$) and $Pmc2_1$ ($[010]$) in $c$-$ePbnm$ (strain plane $[001]$) and $P2_1$ ($[001]$) and $Pm$ ($[110]$) in $ab$-$ePbnm$ (strain plane $[110]$). Of the four polar structures, the $Pmn2_1$ structure is the lowest in energy, which is also consistent with previous work. 

At compressive strains, three polar phases were observed in our calculations. In both the $Pm$([110]) and $Pna2_1$([001]) phases, the direction of the polarization is out of the matching plane. In contrast, in the $P2_1$([001]) phase it is in-plane, which is unexpected for compressive strain in perovskites, in which polarization-strain coupling generally favors polarization along the elongation direction of the unit cell. This suggests that in-plane polarization in $P2_1$ is driven by a different mechanism, which we now discuss.

We performed mode decomposition for the various computed structures at compressive strain using ISOTROPY~\cite{isotropy}. The dominant distortion modes in both the $c$-$ePbnm$($Pbnm$) and $ab$-$ePbnm$($P2_1/m$) structures are $R_4^+$, $M_3^+$ and $X_5^+$, reflecting their relationship to the bulk $Pbnm$ structure. $R_4^+$($M_3^+$) is the out-of-phase (in-phase) oxygen octahedron rotation mode. $X_5^+$ is the displacement of Ca and O in $xy$ plane alternating along $z$-axis.  In the polar structures, besides the zone-center mode $\Gamma_{4z}^-$, which produces the polarization, many other nonpolar modes are also introduced by the symmetry lowering, but most of their amplitudes are very small. Table~\ref{tab:modeDecomposition} shows the dominant modes in the mode decomposition of various structures with $ab$-orientation at 5\% compressive strain. The $Pm$ phase has a large amplitude for the polar mode $\Gamma_{4xy}^-$, as expected for a proper ferroelectric. In contrast, in $P2_1$, the largest additional mode amplitude is not $\Gamma_{4z}^-$, but $X_5^-$, the alternating displacements of Ti and O atoms, and the amplitude of $\Gamma_{4z}^-$ is much smaller than that in other polar phases. Moreover, the eigenvector of $\Gamma_{4z}^-$ in $P2_1$ is quite different from that in $Pm$. At 5\% compressive strain, in $P2_1$, [Ca,Ti,O$_\parallel$,O$_\perp$]=[-0.76,0.19,0.13,0.60], in $Pm$, [Ca,Ti,O$_\parallel$,O$_\perp$]=[-0.10,-0.65,0.55,0.51]. In $P2_1$, the polarization is $A$-site dominated, while in $Pm$ it is $B$-site dominated, as in the other epitaxial-strain-induced ferroelectric phases. The difference between eigenvectors is additional evidence of the different mechanism for the in-plane polarization in $P2_1$.

\begin{figure}[tpd]
\centering
\includegraphics[width=4cm]{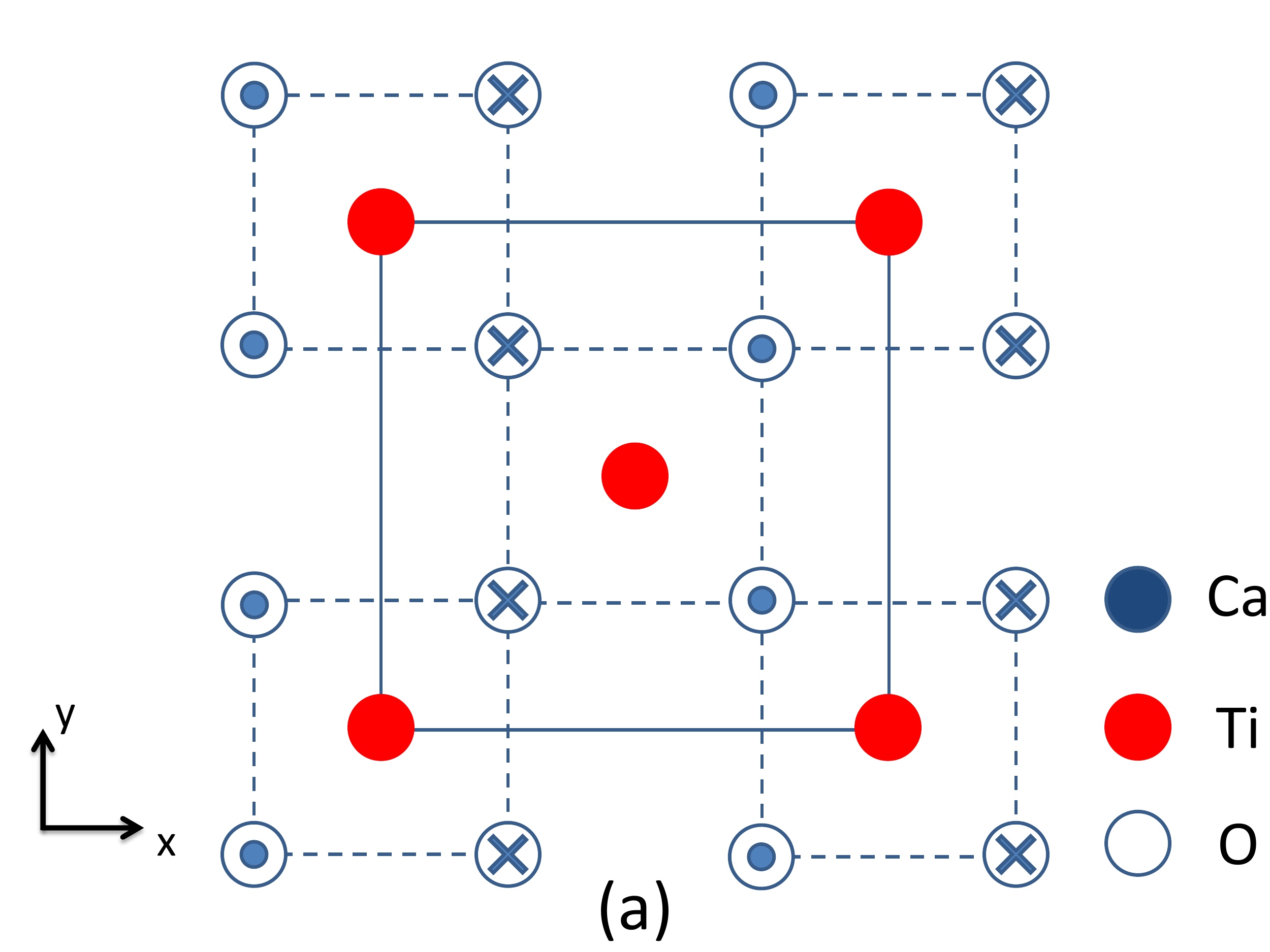}
\includegraphics[width=4cm]{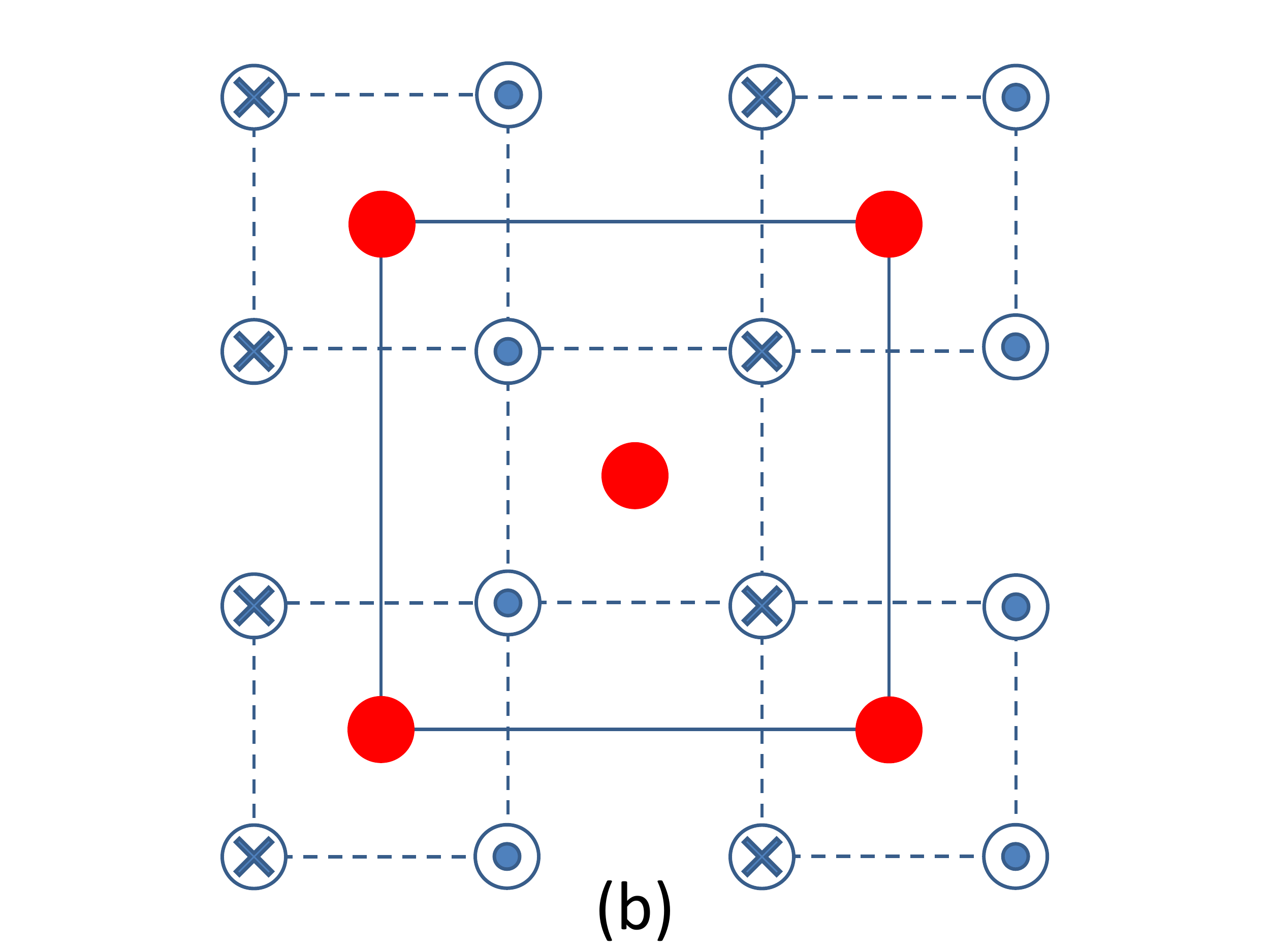}
\includegraphics[width=4cm]{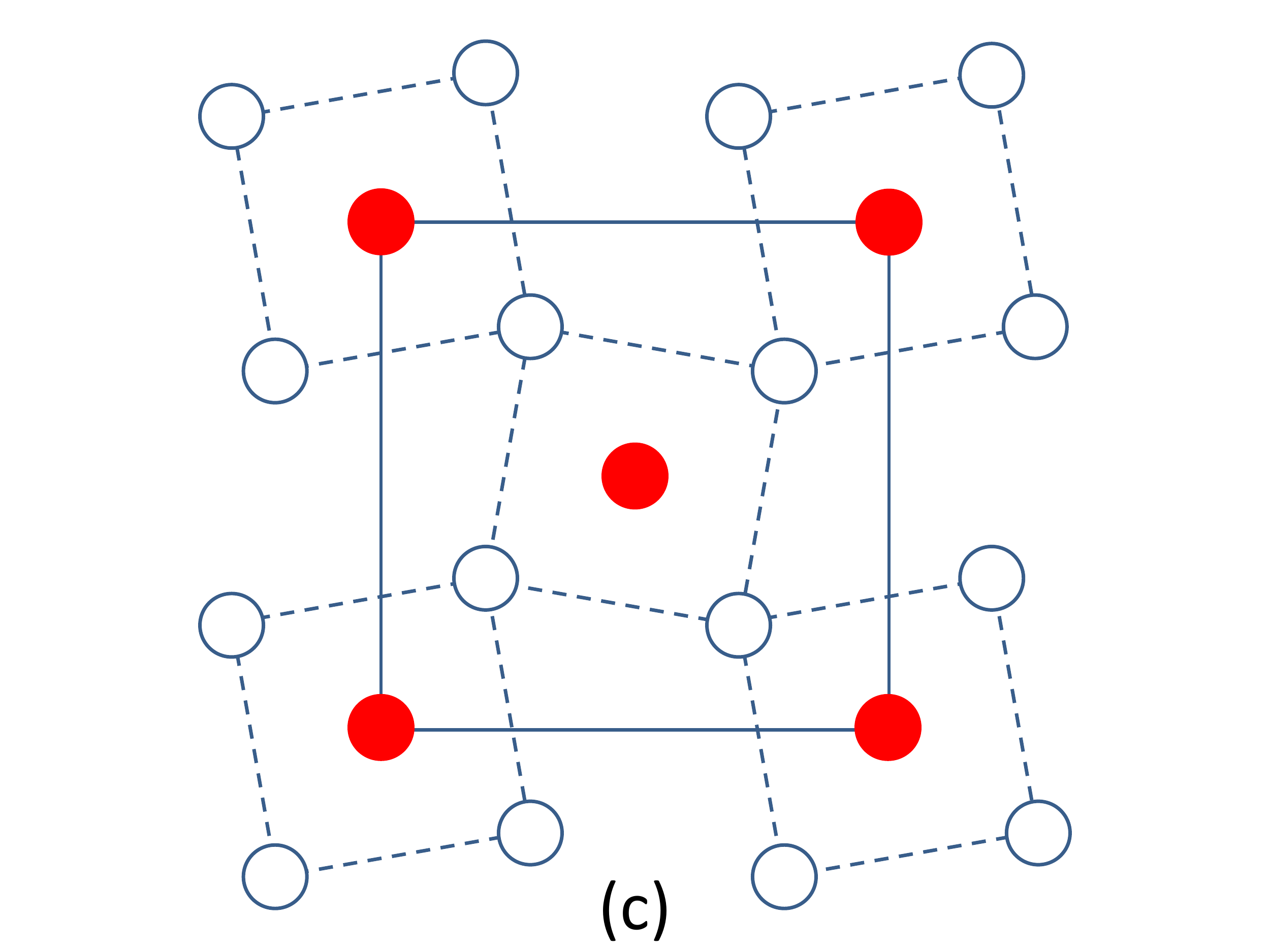}
\includegraphics[width=4cm]{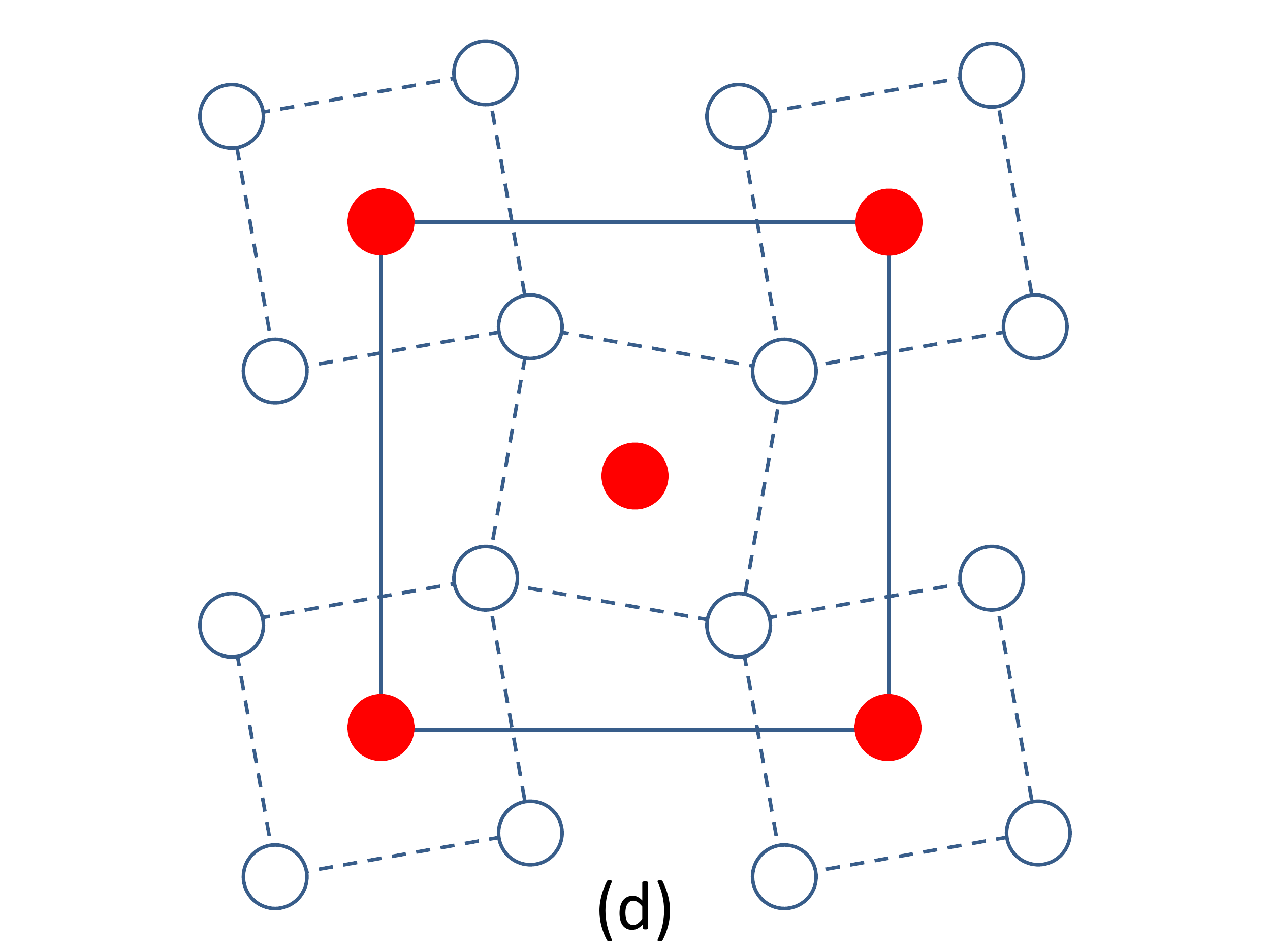}
\includegraphics[width=4cm]{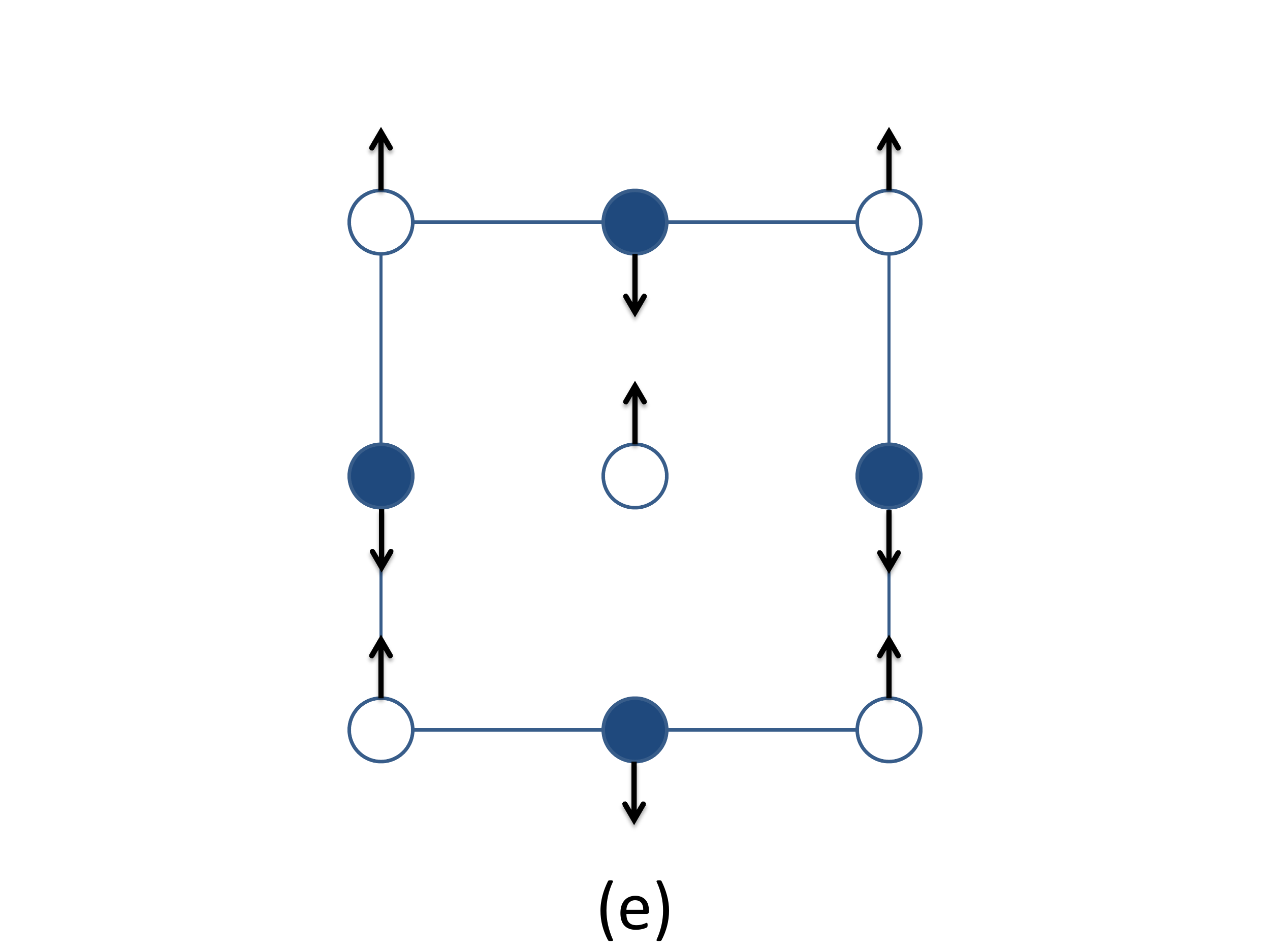}
\includegraphics[width=4cm]{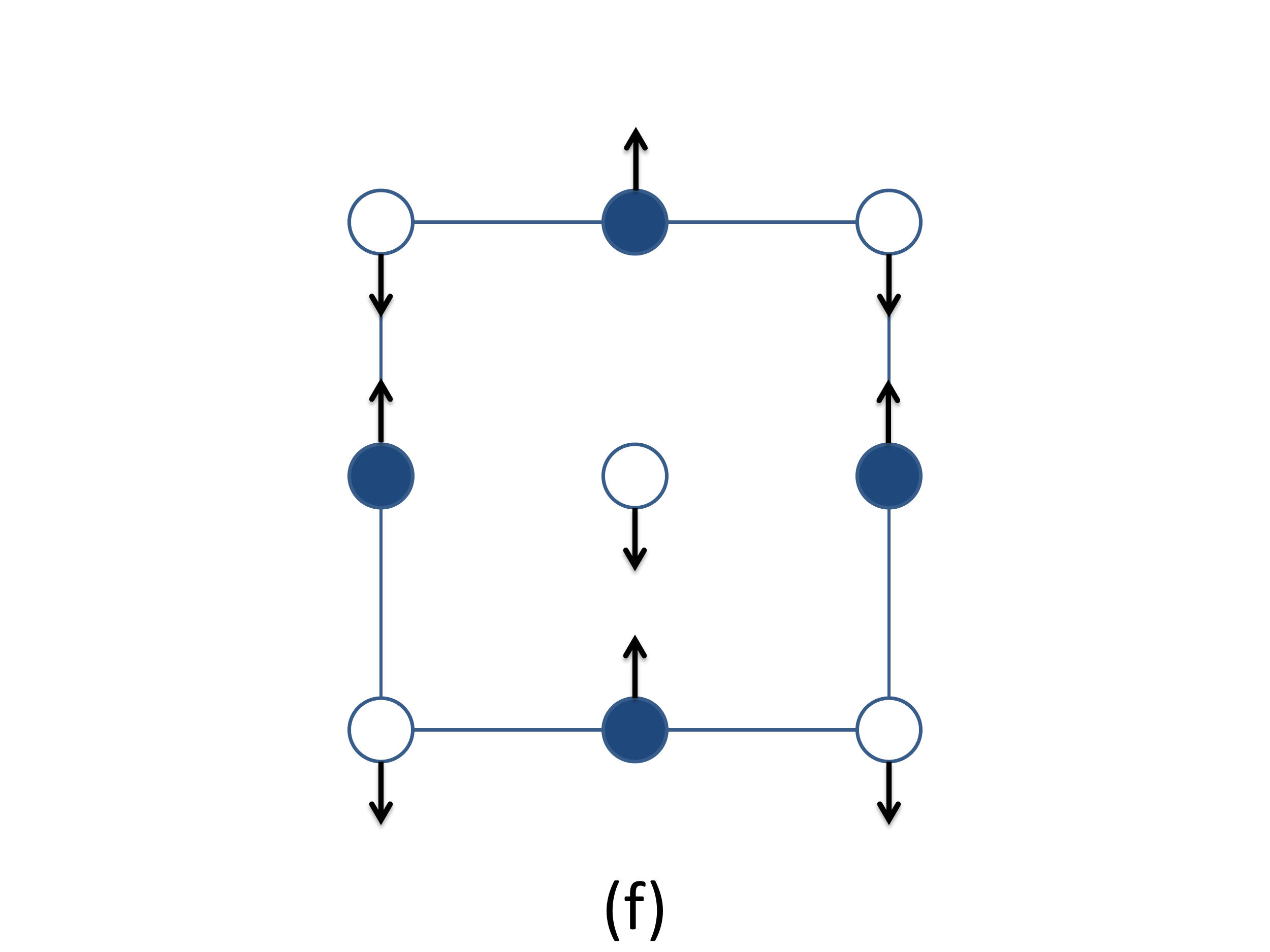}
\includegraphics[width=4cm]{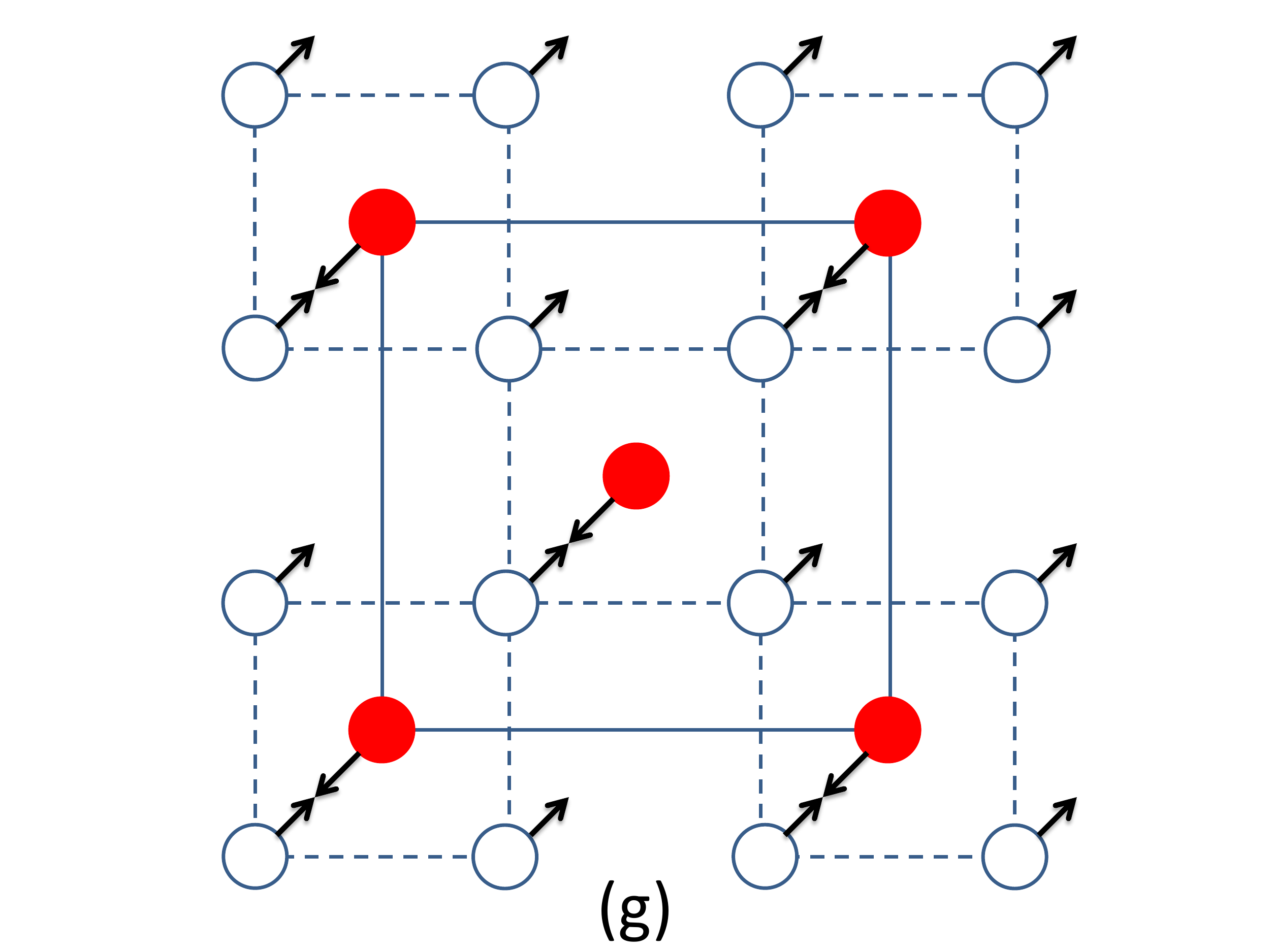}
\includegraphics[width=4cm]{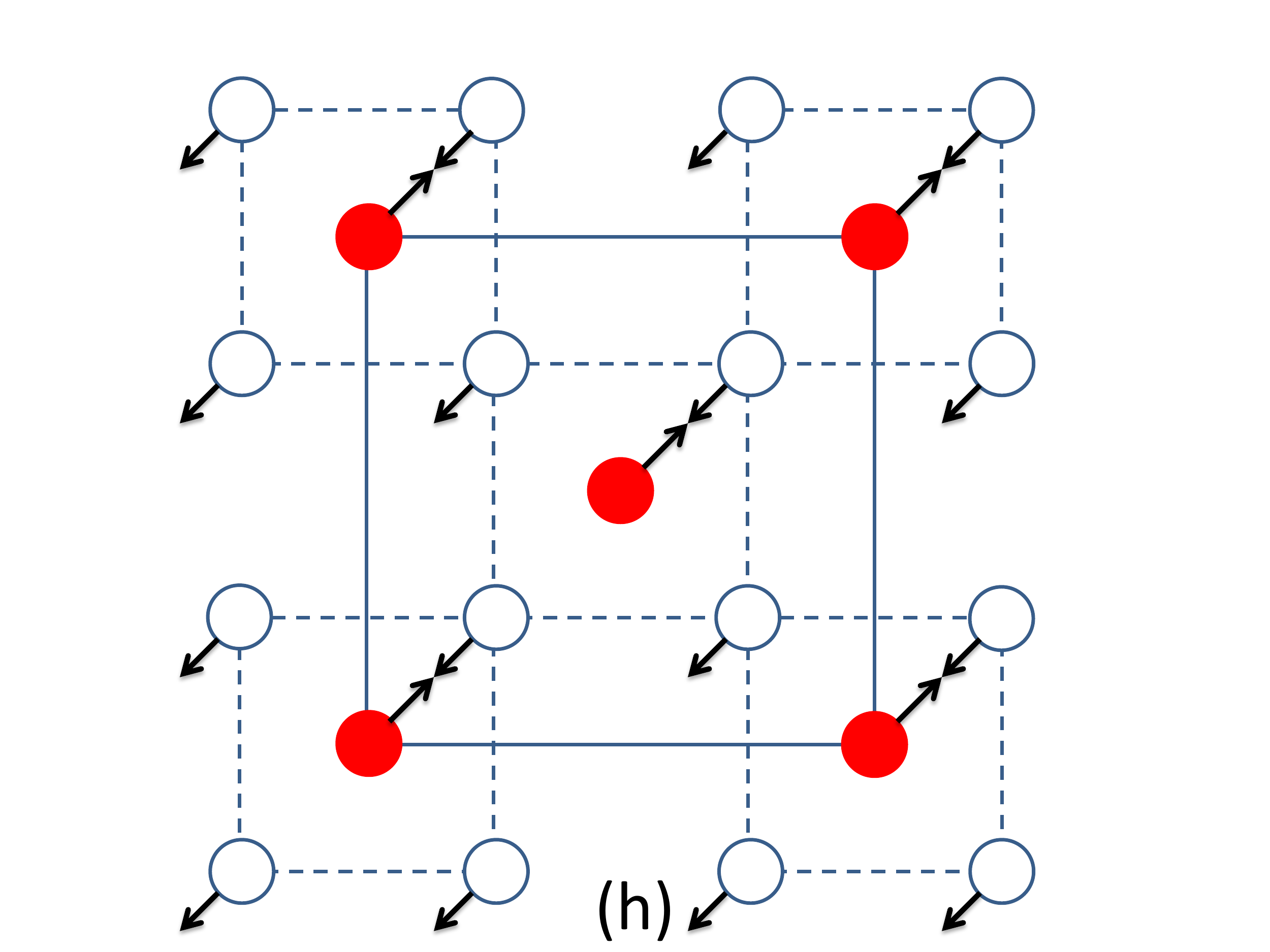}
\caption{Schemes of distortion modes. (a), (b) $R_4^+$ mode in Ti-O layers  at $z=0.0$ and $0.5c$. (c), (d) $M_3^+$ mode in Ti-O layers  at $z=0.0$ and $0.5c$. (e), (f) $X_5^+$ mode in Ca-O layers  at $z=0.25c$ and $0.75c$. (g), (h) $X_5^-$ mode in Ti-O layers  at $z=0.0$ and $0.5c$.}
\label{fig:Mode_disp}
\end{figure}

\begin{table}[h]
\setlength{\tabcolsep}{10pt}
\begin{tabular}{|c|*{3}{p{1.5cm}}|}
\hline
\hline
Mode&$ab$-$ePbnm$&$Pm$&$P2_1$\\
\hline
$R_4^+$&0.777&0.754&0.770\\
$M_3^+$&0.398&0.447&0.430\\
$X_5^+$&0.286&0.302&0.305\\
$\Gamma_4^-$&0&0.160&0.048\\
$X_5^-$&0&0&0.142\\
\hline
\hline
\end{tabular}
\caption{Amplitudes of selected modes in different structures at 5\% compressive strain. 1 a.u. of distortion mode is defined as that the norm of the corresponding displacement eigenvector is 1.00 \AA.}
\label{tab:modeDecomposition}
\end{table}

\begin{figure}
\includegraphics[width=3.0 in]{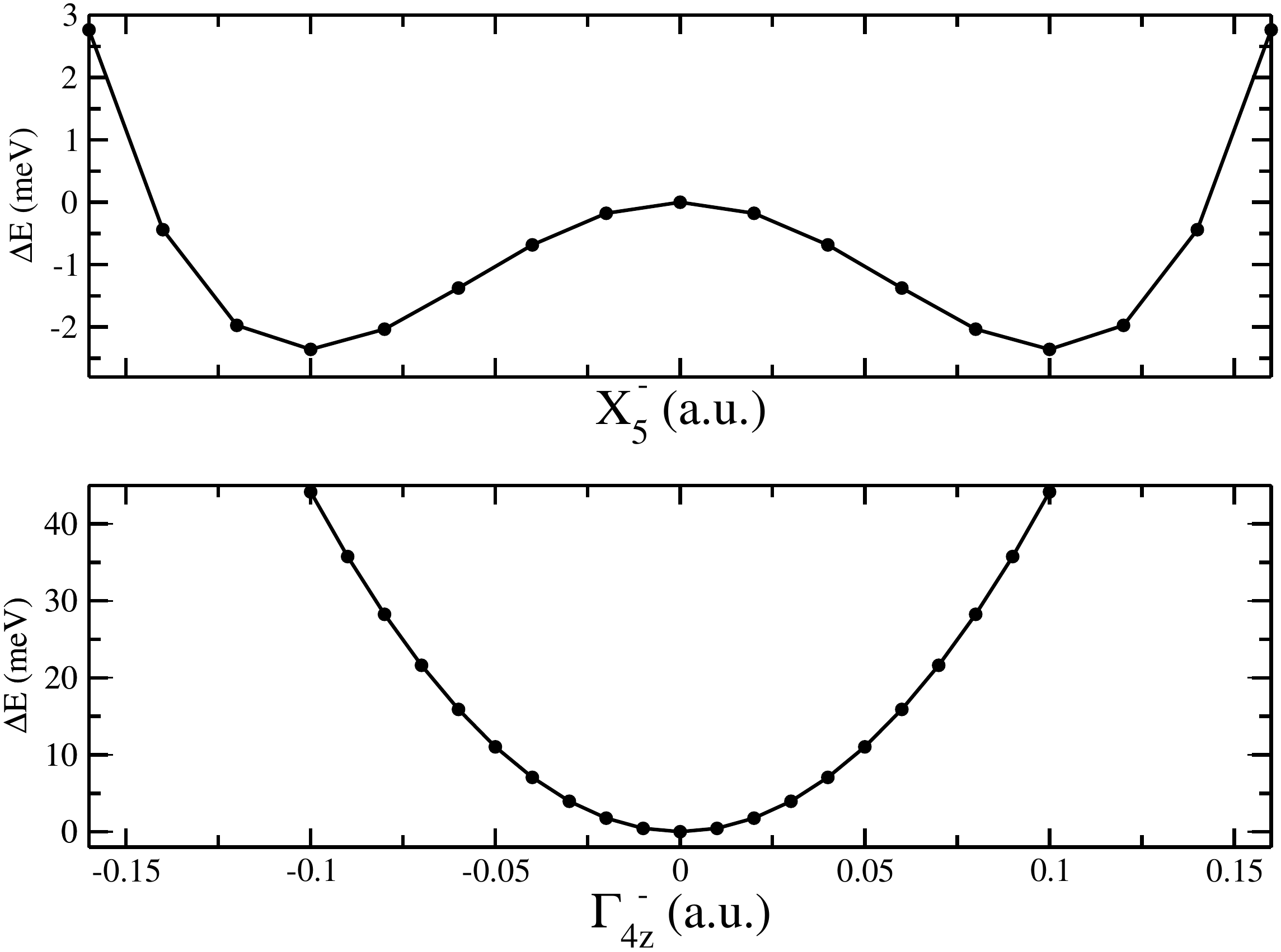}
\caption{Energy as a function of $X_5^-$ and $\Gamma_{4z}^-$. The reference structure is $ab$-$ePbnm$ at 5\% compressive strain.}
\label{fig:gx}
\includegraphics[width=3.0 in]{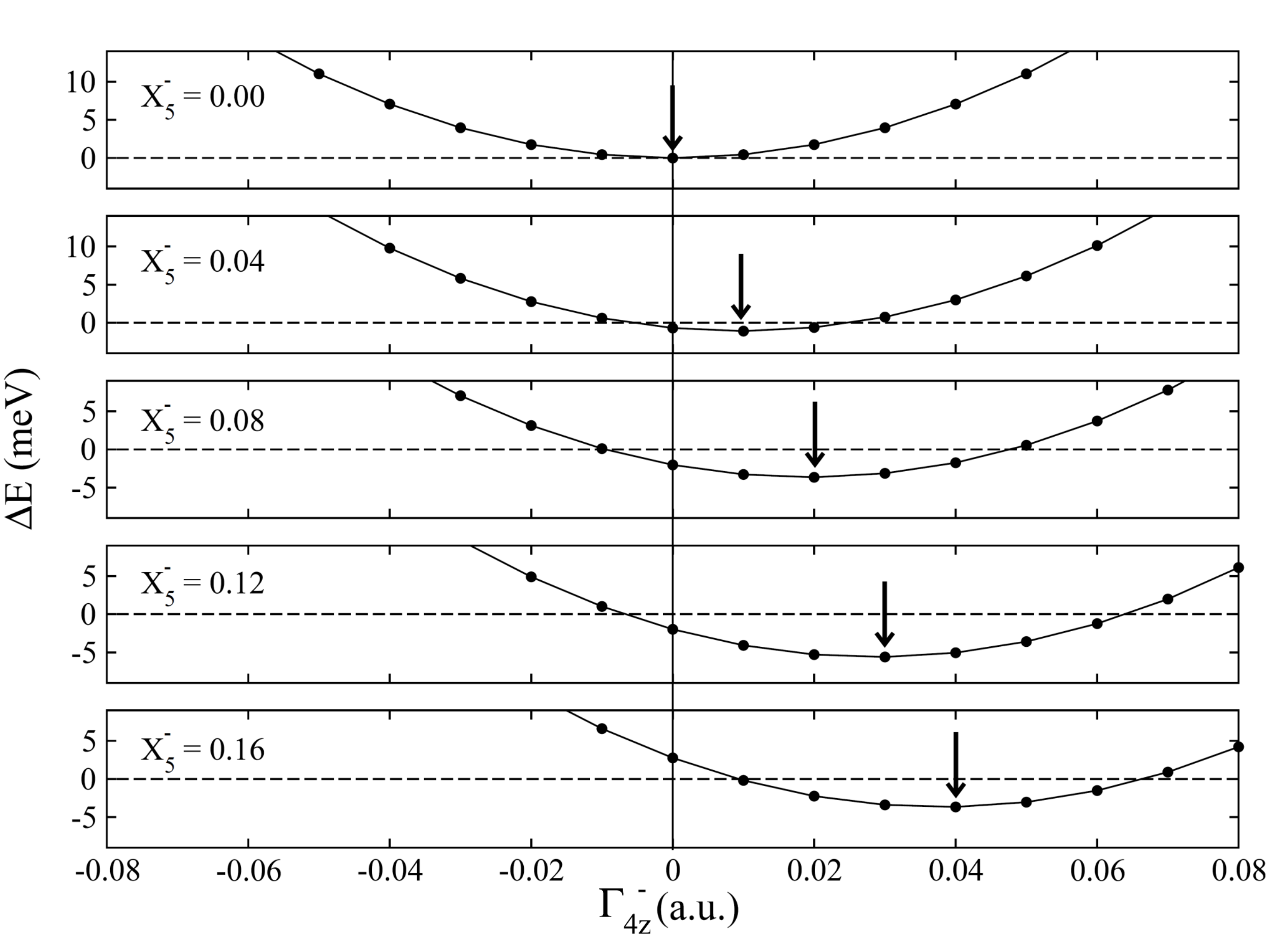}
\caption{Energy as a function of $\Gamma_{4z}^-$ at fixed $X_5^-$.}
\label{fig:gmx}
\end{figure}

To investigate the origin of the polarization in $P2_1$,  we adopted the approach in the YMnO$_3$ study of Fennie and Rabe~\cite{PhysRevB.72.100103}. As shown in Fig.~\ref{fig:gx}, we plotted the energy functions of $X_5^-$ and $\Gamma_{4z}^-$ with $ab$-$ePbnm$ at 5\% compressive strain as the reference structure. We can see that in the reference structure, $X_5^-$ is unstable while $\Gamma_{4z}^-$ is stable, which indicates that the $X_5^-$ mode drives the polar-nonpolar phase transition. Fig.~\ref{fig:gmx} shows that with increasing $X_5^-$ mode amplitude, the energy curve of the $\Gamma_{4z}^-$ mode remains a single well, but the position of its minimum shifts from zero to a non-zero position. This means that the coupling with $X_5^-$ induces the polar $\Gamma_{4z}^-$ mode in the $ab$-$ePbnm$ structure. 
As we will see below, this behavior, characteristic of improper ferroelectrics such as YMnO$_3$, relies on the nonzero amplitude of the distortions in the reference structure, and thus we identify it as hybrid improper ferroelectricity.

Through symmetry analysis, we confirm the origin of the polarization in $P2_1$ as hybrid improper ferroelectricity. Now using the high-symmetry cubic structure as the reference structure, we expand the total energy function $E$ as a 4th-order polynomial in distortion modes $R_4^+, M_3^+,X_5^+,\Gamma_4^-$ and $X_5^-$ (see Figure~\ref{fig:Mode_disp}) and elastic strains $\eta_1, \eta_2$ and $\eta_3$:
\begin{eqnarray*}\label{eq:1}
E&=& E_{Pbnm} (R,M,X,\eta_1,\eta_2,\eta_3)\\
&+& {\kappa} (\Gamma_x^2+\Gamma_y^2+\Gamma_z^2)+{\alpha} (\Gamma_x^2+\Gamma_y^2+\Gamma_z^2)^2 \\
&+&{\lambda}(\Gamma_y^2\Gamma_z^2+\Gamma_x^2\Gamma_y^2+\Gamma_z^2\Gamma_x^2)+\kappa_5X_5^2+\alpha_5 X_5^4 \\
&+& \overline{B}_{x}(\Gamma_x^2+\Gamma_y^2)R^2+\overline{B}_{z}\Gamma_z^2R^2
+B^{\prime}_x(\Gamma_x^2+\Gamma_y^2)M^2 \\
&+&B^{\prime}_z\Gamma^2_zM^2+\tilde{B}_x(\Gamma_x^2+\Gamma_y^2)X^2+\tilde{B}_z\Gamma_z^2X^2\\
&+& \overline{B}_5X_5^2R^2
+B^{\prime}_5X_5^2M^2+ \tilde{B}_5 X_5^2X^2 \\
&+& B_{5x} (\Gamma_x^2+\Gamma_y^2) X_5^2+B_{5z}\Gamma_z^2 X_5^2
+B_{15}(\eta_1+\eta_2)X_5^2 \\
&+&B_{35}\eta_3X_5^2+B_{1x}(\eta_1\Gamma_x^2+\eta_2\Gamma_y^2+\eta_3\Gamma_z^2) \\
& +& B_{1z}((\eta_2+\eta_3)\Gamma_x^2+(\eta_1+\eta_3)\Gamma_y^2+(\eta_1+\eta_2)\Gamma_z^2)\\
&+& C_1 XX_5\Gamma_{z}+C_2RMX_5\Gamma_{z}
\end{eqnarray*}

For simplicity, the sub- and superscripts are omitted when there is no ambiguity; to distinguish the two $X$ modes, $X$ is used for $X_5^+$ and $X_5$ is used for $X_5^-$. To keep the energy polynomial concise, only terms involving $\Gamma_4^-$ and $X_5^-$ are shown explicitly and a single term $E_{Pbnm}$ represents the energy function of non-polar structures. The determination of coefficients from first-principles results~\cite{QZhou_KRabe_CTO} shows that all biquadratic terms, such as $\Gamma_{4z}^{-2}X_5^{-2}$, have positive coefficients, which indicates the mutual competition between any pair of distortion modes. We see that the coupling between $\Gamma_{4z}^-$ and $X_5^-$ originates from the higher order terms $X_5^+X_5^-\Gamma_{4z}^-$ and $R_4^+M_3^+X_5^-\Gamma_{4z}^-$, which is the signature of hybrid improper ferroelectricity. 
The coefficients of $X_5^+X_5^-\Gamma_{4z}^-$ and $R_4^+M_3^+X_5^-\Gamma_{4z}^-$ are 0.46 eV/a.u. and 1.93 eV/a.u.. In the $Pbnm$ phase, $R_4^+$, $M_3^+$ and $X_5^+$ are all nonzero. According to Table~\ref{tab:modeDecomposition}, the product $R_4^+M_3^+$ has an amplitude (0.33 a.u.) similar to that of $X_5^+$ (0.31 a.u.), so that the contributions of the two terms to the induced polarization are comparable. 

This analysis provides a natural way of understanding the instability of the epitaxially-strained nonpolar $Pbnm$ phase to the polar $P2_1$ phase observed in first-principles identification of low-symmetry phases described above. 
The unstable distortion is a mixture of a nonpolar $X_5^-$ mode and a polar $\Gamma_{4z}^-$ mode of the cubic reference structure; we see from the discussion that this coupling results from the pre-existing $R_4^+$, $M_3^+$ and $X_5^+$ distortions in the $Pbnm$ phase. While a transition from the $Pbnm$ phase to the $P2_1$ phase would be accompanied by a divergence in the dielectric constant, following the study of Tol\'edano~\cite{PhysRevB.79.094416} this would be characterized as a pseudoproper, rather than a conventional proper, ferroelectric transition.
%

In hybrid improper ferroelectrics, the multilinear coupling term couples the switching of the zone-center and zone-boundary modes. In the $P2_1$ phase, switching of polarization would result in switching of either  a single mode $X_5^-$ or both $X_5^+$ and the product $R_4^+M_3^+$ . Switching of a rotation mode has an estimated energy barrier of several hundred meV. Fig.~\ref{fig:gx} shows that the well depth of the $X_5^-$ mode is only 2.0 meV, so the $X_5^-$ switching path would be preferred. However, detection of the switching of $X_5^-$ is difficult. It would be possible if there was magnetic order in the system that coupled to $X_5^-$, suggesting further study of other compounds with magnetic order.

In CaTiO$_3$, the improper ferroelectric phase is not the ground state at any value of epitaxial strain in our "strained bulk" calculation.
However, for out-of-plane polarized thin films there are depolarization field effects, which in general would be expected to produce domain structures \cite{fong2004ferroelectricity}, at a cost of free energy increasing with domain wall density. Since for the large polarizations predicted for the out-of-plane polar phases in compressively strained CaTiO$_3$ the domain wall density should be rather high, it might be that the in-plane polarized $P2_1$ phase in ultrathin perovskite films is lower in free energy for some range of strain.  
In addition, the set of perovskites exhibiting the $Pbnm$ structure is quite large, and it might be possible to find a system in which the hybrid improper ferroelectric phase is in fact the ground state for some range of epitaxial strain; a first-principles search is currently in progress. 


In conclusion, our first-principles study of epitaxially strained CaTiO$_3$ revealed the existence of an unexpected polar phase with in-plane polarization at compressive strain. Detailed symmetry analysis indicated that the unusual polarization is due to hybrid improper ferroelectricity and is induced by a strain-induced nonpolar instability at the X point combined with distortion modes present in the nonpolar orthorhombic structure.
This discovery of a novel strain-induced ferroelectric phase in the simple perovskite CaTiO$_3$ provides a new mechanism for the design of functional materials with improper ferroelectricity. 

We thank N. A. Benedek, J. W. Bennett, C. J. Fennie, K. F. Garrity,  D. R. Hamann, J. Hong and D. Vanderbilt for valuable discussions. This work was supported by NSF MRSEC DMR-0820404 and ONR Grant N00014-12-1-1040.

\bibliography{IFbib}
\end{document}